\documentclass[aps,prd,showpacs,twocolumn,10pt,superscriptaddress,floatfix,nofootinbib,longbibliography]{revtex4-1}  

\usepackage{amssymb,amsfonts,amsmath,bm,bbm}   
\usepackage{graphicx}
\usepackage{dcolumn}
\usepackage[
colorlinks=true,
linkcolor=blue,  
breaklinks=true,
urlcolor=blue,
citecolor=blue]{hyperref}
\usepackage{bbold}
\usepackage{epstopdf}

\def\bea{\begin{eqnarray}}
\def\eea{\end{eqnarray}}
\newcommand{\HNUST}{\affiliation{
Hunan Provincial Key Laboratory of Intelligent Sensors and Advanced Sensor Materials, \\ School of Physics and Electronics, Hunan University of Science and Technology, Xiangtan 
 411201, China}} 
      
\newcommand{\UCAS}{\affiliation{
School of Nuclear Science and Technology,
      University of Chinese Academy of Sciences,
      Beijing 100049, China}}
      
\newcommand{\IHEP}{\affiliation{
Institute of High Energy Physics, Chinese Academy of Sciences, Beijing 100049, China}}

\newcommand{\Squ}{\affiliation{
College of Information Engineering, Suqian University, Suqian 223800, China}}

\begin{document}

\preprint{APS/123-QED}


\title{Cold quark matter in a quasiparticle model: thermodynamic consistency and stellar properties}

\author{Zhi-Jun Ma} 
\HNUST

\author{Zhen-Yan Lu} \email{luzhenyan@hnust.edu.cn}\HNUST

\author{Jian-Feng Xu} 
\Squ

\author{Guang-Xiong Peng} 
\UCAS\IHEP

\author{Xiangyun Fu} 
\HNUST

\author{Junnian Wang}
\HNUST

\date{\today}

\begin{abstract}
The strong coupling in the effective quark mass was usually taken as a constant in a quasiparticle model while it is, in fact, running with an energy scale. With a running coupling, however, the thermodynamic inconsistency problem appears in the conventional treatment. We show that the renormalization subtraction point should be taken as a  function of the summation of the biquadratic chemical potentials if the quark's current masses vanish, in order to ensure full thermodynamic consistency. Taking the simplest form, we study the properties of up-down ($ud$) quark matter, and confirm that the revised quasiparticle model fulfills the quantitative criteria for thermodynamic consistency.  
Moreover, we
find that the maximum mass of an $ud$ quark star 
can be larger than two times the solar mass, reaching up to $2.31M_{\odot}$, for reasonable model parameters. 
However, to further satisfy the upper limit of tidal deformability $\tilde{\Lambda}_{1.4}\leq 580$ observed in the event GW170817, the maximum mass of an $ud$ quark star 
can only be as large as $2.08M_{\odot}$, namely 
$M_{\text{max}}\lesssim2.08M_{\odot}$. 
In other words, our results indicate that the measured tidal deformability for event GW170817 places an upper bound on the maximum mass of $ud$ quark stars, but which does not rule out the possibility of the existence of quark stars composed of $ud$  
quark matter, with a mass of about two times the solar mass.

\end{abstract}

\maketitle

\section{Introduction} \label{sec:Introduction}

The accepted theory for a precise description of the strong interaction between quarks and gluons is the Quantum Chromodynamics (QCD)~\cite{Gross-2022hyw}, in which the running coupling
$\alpha_{s}(\Lambda^2)$ acts as one of the basic parameters. The relevant theoretical and experimental researches show that QCD has two special features: color confinement and asymptotic freedom \cite{Gross-1973ju, Gross-1973id,Politzer-1973fx}. The former has been proposed 
because a single quark cannot be observed from experiments even at the present time,  but the quantum number of the so-called color is indispensable in order to explain the confusion of the requisite antisymmetry wave function of hadrons. The latter reveals that the interaction between quarks gets weaker with increasing energy scales. These two extraordinary features of QCD happen at two extreme energy scale: the asymptotic freedom is related to the high-energy region while the color confinement comes up in the infrared domain. 
Due to the difficulties in the non-perturbative regime and  the consistent implementation of finite density in lattice simulations, some QCD-inspired phenomenological models 
have been constructed to
study the properties of strongly interacting quark matter, 
e.g. the MIT bag model \cite{Farhi-1984qu, Chakrabarty-1996te,Weber-2004kj,74Chodos.Jaffe.ea3471-3495PRD}, Nambu-Jona-Lasinio (NJL) model \cite{Rehberg-1995kh,Kunihiro-1989my,Hanauske-2001nc,PeresMenezes-2005bc,Menezes-2008qt,Menezes-2009uc}, 
perturbative QCD model \cite{Fraga-2004gz,Peng-2005xp,Kurkela-2009gj}, density-dependent quark mass model~\cite{Peng-1999gh,Wen-2005uf,Lu-2016jsv,Chu-2012rd,Backes-2020fyw}, etc.

Employing a two-flavor 
NJL model including the vector interactions, 
Yuan $et~al.$~\cite{Yuan-2022dxb} found that there is a possibility for both $ud$ and strange ($uds$) quark matter being absolutely stable, depending on the interplay of the confinement with quark vector interaction and the exchange interaction channels. As stated by Buballa in Ref.~\cite{05Buballa-PhysRep}, however,  
the NJL model does not support the idea
that 
$uds$ quark matter is more stable than the $ud$ quark matter. 
In particular, by using an effective theory with Yukawa coupling to quarks Holdom $et~al.$~\cite{Holdom-2017gdc} demonstrated that $ud$ quark matter generally has lower bulk energy per
baryon than normal nuclei and $uds$ quark matter. 
Moreover, within a confining quark matter model, Cao $et~al.$~\cite{Cao-2020zxi} demonstrated that the reported
GW190814's secondary component with a mass 
greater than two times the solar mass 
could be an $ud$ quark star.  
After the proposal that $ud$ 
quark matter is more stable than $uds$ quark matter, the study of the properties of strongly interacting $ud$ quark matter and its astrophysical implications has recently received
increased interest in the literature, e.g.,  Refs.~\cite{Zhang-2020jmb,Zhao-2019xqy,Wang-2019jze,Ren-2020tll,Li-2022vof,Restrepo-2022wqn,Xu-2021alh}.

The quasiparticle model, as one of the QCD-inspired phenomenological models, has also been extensively studied and widely applied to the study of quark matter over the last two 
decades. 
In the case of a hot gluon gas, the equation of state with the effective mass agrees well with the lattice results even at temperatures where the QCD coupling constant is not small \cite{ Peshier-2002ww,Peshier-1995ty,
Schertler-1996tq, Schertler-1997vv}. Motivated by the successful application of the hard-thermal-loop approximation at finite temperature, many authors have applied 
the hard-dense-loop 
approximation at finite density to investigate the properties of the strongly interacting cold quark matter \cite{Schertler-1996tq, Schertler-1997vv, Wen-2009zza,Wen-2012jw}. 
Its original version keeps all the thermodynamic quantities
in the same form as those of a free particle system~\cite{Biro-1990vj,Goloviznin-1992ws,Peshier-1994zf}. The thermodynamic treatment in this model has been much discussed~\cite{Gorenstein-1995vm,Peshier-1999ww,Bannur-2005wm}. 
If the QCD coupling 
is treated as a real constant, the
thermodynamic inconsistency problem can be removed by
including an additional term in the thermodynamic potential density
\cite{Wen-2012jw}. 
However, the actual QCD coupling is running, i.e. it depends on 
where the subtraction point is chosen in the renormalization scheme~\cite{Prosperi-2006hx,Deur-2016tte}. 
In high energy physics, the subtraction point $\Lambda$ is usually taken as
the momentum transfer of the corresponding process.
While in dense quark matter,
it should be a function of the chemical potentials of the quarks
through the subtraction point
~\cite{Fraga-2005nu,Ipp-2003yz,Fraga-2001id}. 
In this case, the additional term is 
determined by
a path integral. Because the thermodynamic quantity
is a state function, it should thus be independent of the chosen path.
Otherwise, the thermodynamic inconsistency problem will appear~\cite{Peng-1999gh,Xia-2014zaa}:
the pressure at the minimum energy per baryon deviates from zero,
which contradicts fundamental thermodynamics. 
In this work, we will show that the renormalization-group  
subtraction point should be generally taken
as a function of the biquadratic summation of respective chemical potential
for quarks.  

In this study, we focus on the thermodynamic inconsistency problem of a two-flavor quasiparticle model at zero temperature and finite chemical potential with an effective quark mass. 
Within the revised quasiparticle model, we are particularly interested in the stability windows and thermodynamic properties of the $ud$ quark matter. 
Through our research, we seek to understand the behavior of quark matter under extreme conditions and its implications for cosmological and astrophysical phenomena including, in particular quark star structure and tidal deformability. 
The results for the mass-radius relation as well as the tidal deformability of quark stars indicate that the recently observed massive pulsars with a mass of about two solar masses could be made of $ud$ quark matter. 

The rest of the paper is organized as follows. In Sec.~\ref{sec:modelI}, we review the conventional treatment in the quasiparticle model with an effective quark mass and show the
thermodynamic inconsistency in this treatment. Then in Sec.~\ref{sec:modelII}, we limit the form of the renormalization subtraction point as a function of the chemical potentials and  determine a medium-dependent effective bag constant which
makes the thermodynamic treatment self-consistent.   
In Sec.~\ref{sec:results}, we present the numerical results for the calculations. 
Finally, a summary is given in Sec.~\ref{sec:conclusion}.

\section{Thermodynamic inconsistency in the conventional version of quasiparticle model}  \label{sec:modelI}

Due to the strong interaction between quarks, the quark masses are medium-dependent, namely, the quark masses vary with the environment. 
The effective quark mass is obtained from the zero momentum limit of the quark dispersion relation, which follows the so-called hard-dense-loop approximation of the quark self-energy, giving~
\cite{Schertler-1996tq,Srivastava-2010xa}
\begin{eqnarray}\label{m_(i)}
m^*_{i}=\sqrt{\frac{2\alpha_{s}}{3\pi}}\mu_{i}~~~~~~~~(i=\mathrm{u},~\mathrm{d}),
\end{eqnarray}
where $\mu_{i}$  
 represents the chemical potential for the $i$th flavor of quarks. 

At zero temperature, the quasiparticle contribution to the thermodynamic potential density of the system is given by 
\begin{eqnarray}\label{eq:Omegai}
\Omega_{i}
&=&
-\frac{g_{i}}{48\pi^2}\left(\mu_{i}k_{F,i}(2\mu_{i}^2-5{m^{*}_{i}}^2)
+
3{m_{i}^{*}}^4\ln\frac{\mu_{i}+k_{F,i}
}{m_{i}^{*}}\right),\nonumber\\
\label{Omega_i}
\end{eqnarray}
where $k_{F,i}=\sqrt{\mu_i^2-m_i^{*2}}$ is the Fermi momentum, and $g_{i}$ is the statistical factor with $g_{i}$=6 for quarks and $g_{i}=2$ for electrons. 
In addition, 
the thermodynamic potential density of 
electrons is $\Omega_{\mathrm{e}}=-\mu_{\mathrm{e}}^4/12\pi^2$ 
with $\mu_{\mathrm{e}}$ being the electron chemical potential.

\begin{figure}[h]  
  \includegraphics[width=0.48\textwidth]{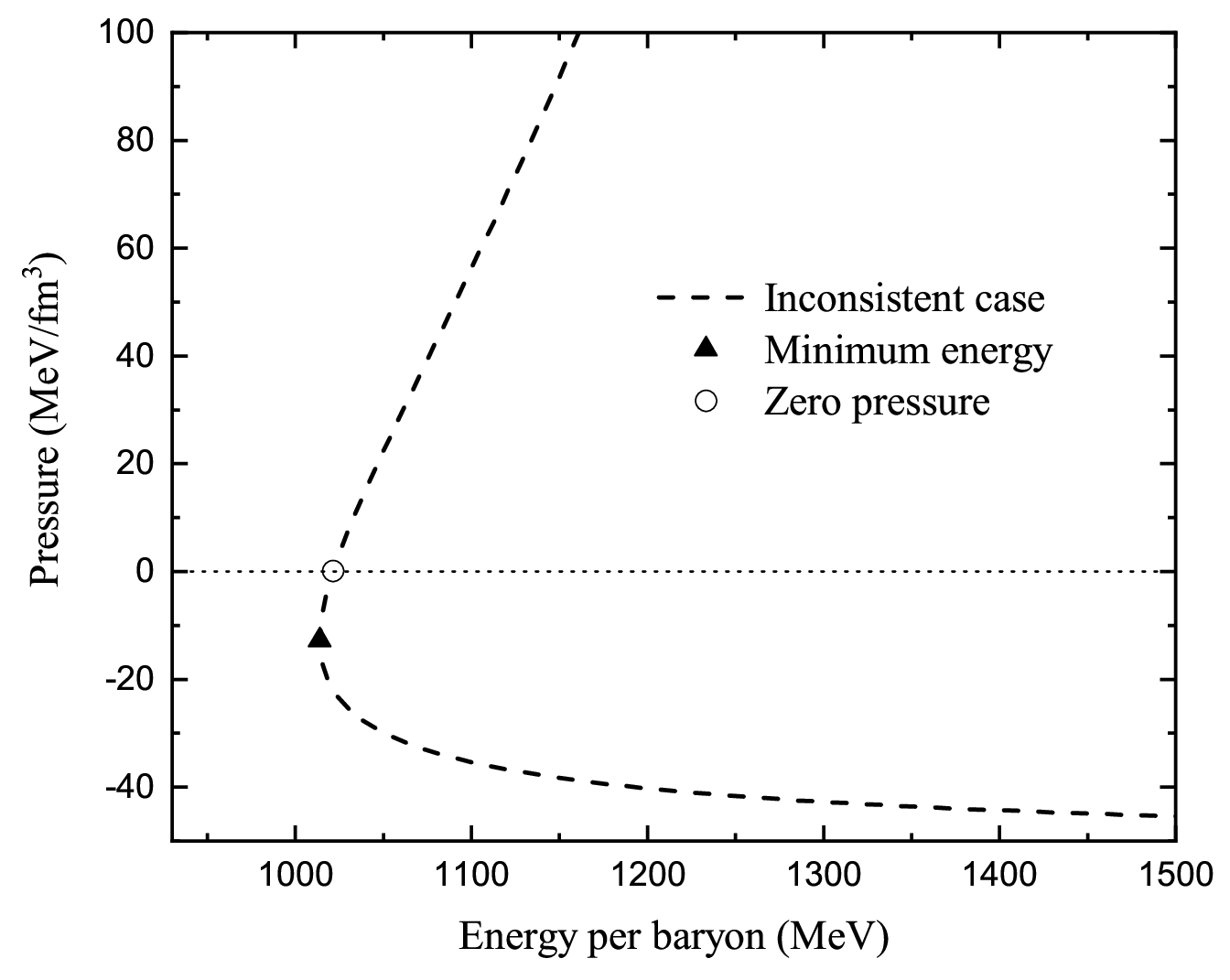}\\
  \caption{Pressure versus energy per baryon of $ud$ quark matter in the quasiparticle model without adding an effective bag constant (inconsistent case). It is clear that the energy minimum denoted by a triangle does not locate at the zero pressure point denoted by a circle.
  }\label{fig:fig1inconsistent}
\end{figure}

By using the basic relations of
standard thermodynamics $P=-\Omega$ and $E=\Omega+\mu n$ with $\Omega$ being the thermodynamic potential density of the system, $n$ the number density of the particles, 
in Fig.~\ref{fig:fig1inconsistent} we plot the energy per baryon of $ud$ quark matter as a function of the pressure without adding an effective bag constant.

As proved in Refs.~\cite{Peng-2000ff,08Peng.Li.ea65807-65807PRC,Xia-2014zaa} that any consistent thermodynamic treatment must
ensure that $\delta=P-n^2\frac{\mathrm{d}}{\mathrm{d}n}\left(\frac{E}{n}\right)=0$ 
at arbitrary density. 
In particular, the pressure at the minimum energy per baryon must be zero. From Fig.~\ref{fig:fig1inconsistent}, however, it is evident that the energy minimum does not locate at the zero pressure point, as required by the thermodynamic consistency of phenomenological models.
In other words, the quasiparticle model with a chemical potential-dependent quark mass but without adding an effective bag constant is thermodynamically inconsistent. 
We should emphasize that 
if one takes
the strong coupling constant $\alpha_s$ as a free input parameter and also adds a medium-dependent effective bag constant to the thermodynamic potential density of the system, as was done in Refs.~\cite{Wen-2009zza,Chu-2022ofc,Chu-2021aaz,Chu-2021oey,Zhang-2021qhl}, the phenomenological models do not encounter thermodynamic inconsistency problems, but we note that the strong coupling runs with the energy scale. 
Hence, in the next section we will show that a thermodynamically determined effective bag constant with a running strong coupling constant can solve this problem properly.

\section{Self-consistent thermodynamic treatment at finite chemical potentials}
\label{sec:modelII}

The quasiparticle model is based on the idea that quarks can be treated as 
quasiparticles with an effective quark mass, which is determined by the quark self-energy and quark-quark interaction. 
The self-consistency of this model 
is maintained by adding a new term, which is a chemical potential dependent variable at zero temperature. In the two-flavor case, the total thermodynamic potential density of the system can be written as \cite{Wen-2009zza}
\begin{eqnarray}\label{eq:Omega}
\Omega
&=&\sum_{i}\Omega_{i}(\mu_{i},m_{i}^*)+B^*(\{\mu_i\}), \label{OMEGA}
\end{eqnarray}
where $B^{*}(\{\mu_i\})\equiv
B(\{\mu_{i}\})
+B_{0}$ is defined as an effective bag constant 
including the energy contributions both from the vacuum $B_{0}$ and medium effects $B(\{\mu_{i}\})$. 
Following the procedure proposed in Refs.~\cite{Lu-2016fki}, 
we have
\begin{eqnarray}\label{eqn:n_u}
n_{i}&=&-\frac{\mathrm{d}\Omega}{\mathrm{d}\mu_{i}}\bigg|_{\mu_{j\neq i}}
=-\frac{\partial\Omega_{i}}{\partial\mu_{i}}\nonumber\\
&&-\underbrace{\Bigg[\frac{\partial\Omega_{i}}{\partial m_{i}^*}\frac{\partial m_{i}^*}{\partial\mu_{i}}
+\sum_{j=\mathrm{u},\mathrm{d}}\frac{\partial\Omega_{j}}{\partial m_{j}^*}\frac{\partial m_{j}^*}{\partial\alpha_{s}}\frac{\partial\alpha_{s}}{\partial \Lambda}\frac{\partial \Lambda}{\partial\mu_{i}}
+\frac{\partial B}{\partial\mu_{i}}\Bigg]}_{=0}~~~
\end{eqnarray}
for $i$th flavor quark in the two-flavor quasiparticle model. The 
derivative of the quasiparticle contribution of the thermodynamic potential density with respect to the effective quark mass $m_i^{*}$ is given by
\begin{eqnarray}
\frac{\partial\Omega_{i}}{\partial m_{i}^*}&=&\frac{g_{i}m_{i}^*}{4\pi^2}\bigg(\mu_{i}k_{F,i}
-{m_{i}^*}^2\ln\frac{\mu_{i}+k_{F,i}
}{m_{i}^*
}\bigg).\label{Omega_diffm_i}
\end{eqnarray}

From the quasiparticle point of view, 
quarks are considered 
quasiparticles with an effective quark mass generated by medium effects. In this case, 
the quark number density should have the same form as in the standard statistical mechanics, with the current quark mass replaced by the effective mass. i.e.,
\begin{eqnarray}
n_{i}
=\frac{g_{i}}{6\pi^2}k_{F,i}^3.
\label{eq:nistandard}
\end{eqnarray}
This is, in fact, 
the first term on the right hand side of the second equality of Eq.~(\ref{eqn:n_u}). 
By also considering the contribution from the down quark, 
we immediately get
\begin{align} \label{eq:dB}
\mathrm{d}B=-\mathcal{F}\mathrm{d}\mu_\mathrm{u}-\mathcal{M}\mathrm{d}\mu_\mathrm{d},
\end{align}
where
\begin{eqnarray}\label{P(mu_u,mu_d)}
\mathcal{F}(\mu_\mathrm{u},\mu_\mathrm{d})&=&\frac{\partial\Omega_\mathrm{u}}{\partial m_\mathrm{u}^*}\frac{\partial m_\mathrm{u}^*}{\partial\mu_\mathrm{u}}
+\sum_{j=\mathrm{u},\mathrm{d}}\frac{\partial\Omega_{j}}{\partial m_{j}^*}\frac{\partial m_{j}^*}{\partial\alpha_{s}}\frac{\partial\alpha_{s}}{\partial \Lambda}\frac{\partial \Lambda}{\partial\mu_\mathrm{u}},~~~~~
\end{eqnarray}

\begin{eqnarray}\label{M(mu_u,mu_d)}
\mathcal{M}(\mu_\mathrm{u},\mu_\mathrm{d})&=&\frac{\partial\Omega_\mathrm{d}}{\partial m_\mathrm{d}^*}\frac{\partial m_\mathrm{d}^*}{\partial\mu_\mathrm{d}}
+\sum_{j=\mathrm{u},\mathrm{d}}\frac{\partial\Omega_{j}}{\partial m_{j}^*}\frac{\partial m_{j}^*}{\partial\alpha_{s}}\frac{\partial\alpha_{s}}{\partial \Lambda}\frac{\partial \Lambda}{\partial\mu_\mathrm{d}}.~~~~~
\end{eqnarray}

Because the thermodynamic
quantity should be a function of the independent state variables, the integral of Eq.~(\ref{eq:dB}) should be path independent~\cite{Xu-2014zea,Xu-2015wya}. 
Mathematically, this can be achieved by imposing the following condition:  
\begin{eqnarray}\label{partialFM}
\frac{\partial \mathcal{F}(\mu_\mathrm{u},\mu_\mathrm{d})}{\partial\mu_\mathrm{d}}=\frac{\partial \mathcal{M}(\mu_\mathrm{u},\mu_\mathrm{d})}{\partial\mu_\mathrm{u}}.
\end{eqnarray} 
Obviously, if the QCD coupling constant was assumed to be a pure constant as in Ref.~\cite{Wen-2009zza}, the above Cauchy conditions in Eq.~(\ref{partialFM}) would always be satisfied. However, the QCD coupling is running with the quark chemical potentials. We thus need to find the condition that ensures the equality in Eq.~(\ref{partialFM}).
Substituting 
the explicit expressions for 
$\mathcal{F}$ and $\mathcal{M}$ in Eqs.~(\ref{P(mu_u,mu_d)}) and (\ref{M(mu_u,mu_d)}) into Eq. (\ref{partialFM}) and simplifying the corresponding expression then gives
\begin{eqnarray}
\mu_\mathrm{u}^3\frac{\partial \Lambda}{\partial\mu_\mathrm{d}}=\mu_\mathrm{d}^3\frac{\partial \Lambda}{\partial\mu_\mathrm{u}}.\label{muU3}
\end{eqnarray}
Eq.~(\ref{muU3}) is a quasilinear partial differential equation, and its general solution is an arbitrary function of $\mu_\mathrm{u}^4+\mu_\mathrm{d}^4$, i.e.,
$
\Lambda=f(\mu_\mathrm{u}^4+\mu_\mathrm{d}^4). 
$ 
Since the subtraction point $\Lambda$ carries the
dimension of an energy, we take the simplest form
\begin{eqnarray}\label{Qexpressin}
\Lambda=C\sqrt[4]{\frac{\mu_\mathrm{u}^4+\mu_\mathrm{d}^4}{N_{\mathrm{f}}}},
\end{eqnarray}
where $C$ is introduced as a dimensionless model parameter, and $N_{\mathrm{f}}=2$ is the number of flavors.

In this case, the bag constant $B$ can be obtained by a path integral as 
\begin{eqnarray}\label{integrationB}
B=\int_{(0,0)}^{(\mu_\mathrm{u},\mu_\mathrm{d})} \mathcal{F}\left(\mu_{\mathrm{u}}^{\prime}, \mu_{\mathrm{d}}^{\prime}\right) \mathrm{d} \mu_{\mathrm{u}}^{\prime}+\mathcal{M}\left(\mu_{\mathrm{u}}^{\prime}, \mu_{\mathrm{d}}^{\prime}\right) \mathrm{d} \mu_{\mathrm{d}}^{\prime}.~~~
\end{eqnarray}

\section{Numerical results and discussions}
\label{sec:results}

Considering a system composed of up, down quarks and electrons 
in $\beta$ equilibrium achieved by the weak reactions $d\leftrightarrow u+e+\bar{\nu}_e$, one easily obtains the following equality
\begin{align}\label{eq:muUDe}
\mu_\mathrm{u}=\mu_\mathrm{d}+\mu_{\mathrm{e}}.
\end{align}
Neutrinos are assumed to enter and escape the system freely, so the chemical potential of neutrinos is zero. The baryon number density $n_{\mathrm{b}}$ is defined as
\begin{align}\label{eq:nB}
n_{\mathrm{b}}=\frac{1}{3}(n_\mathrm{u}+n_\mathrm{d}).
\end{align}
The charge neutrality condition is also imposed on the system, i.e.
\begin{align}\label{eq:nQ0}
\frac{2}{3}n_\mathrm{u}-\frac{1}{3}n_\mathrm{d}-n_{\mathrm{e}}=0.
\end{align}

The strong coupling constant
$\alpha_{s}$, which is related to the running coupling constant $g$  by $\alpha_{s}=g^2/4\pi$, is, in fact, 
a function of the renormalization subtraction point $\Lambda$. 
In this work, we adopt the analytic QCD coupling constant proposed by Shirkov $et~al.$ at the one-loop level~\cite{Shirkov-1997wi}, which is written as
\begin{align}\label{alpha_s}
\alpha_{s}(\Lambda^2)=\frac{4\pi}{\beta_{0}}\bigg[\frac{1}{\ln(\Lambda^2/\Lambda_{\text{QCD}}^2)}+\frac{\Lambda_{\text{QCD}}^2}{\Lambda_{\text{QCD}}^2-\Lambda^2}\bigg],
\end{align}
where $\beta_{0} = 11-2N_{\mathrm{f}}/3$, the QCD scale parameter $\Lambda_{\text{QCD}}=147$ MeV~\cite{Lu-2022khf}, and $N_{\mathrm{f}}=2$ is the quark flavor considered in the system.

For a given baryon density $n_{\mathrm{b}}$, the chemical potentials $\mu_{i}$ are obtained by solving a set of Eqs.~(\ref{eq:muUDe})-(\ref{eq:nQ0}). 
Because the integrand in Eq.~(\ref{integrationB}) is long and complicated, 
only numerical studies will be possible. 
Accordingly, by substituting Eq.~(\ref{integrationB}) into Eq.~(\ref{OMEGA}), the energy density and the pressure of the system can then be derived
\begin{align}\label{eq:E_P}
E=\Omega+\sum_{i}\mu_{i}n_{i},
~~~~~~P=-\Omega.
\end{align}

\begin{figure}[h]
  \includegraphics[width=0.48\textwidth]{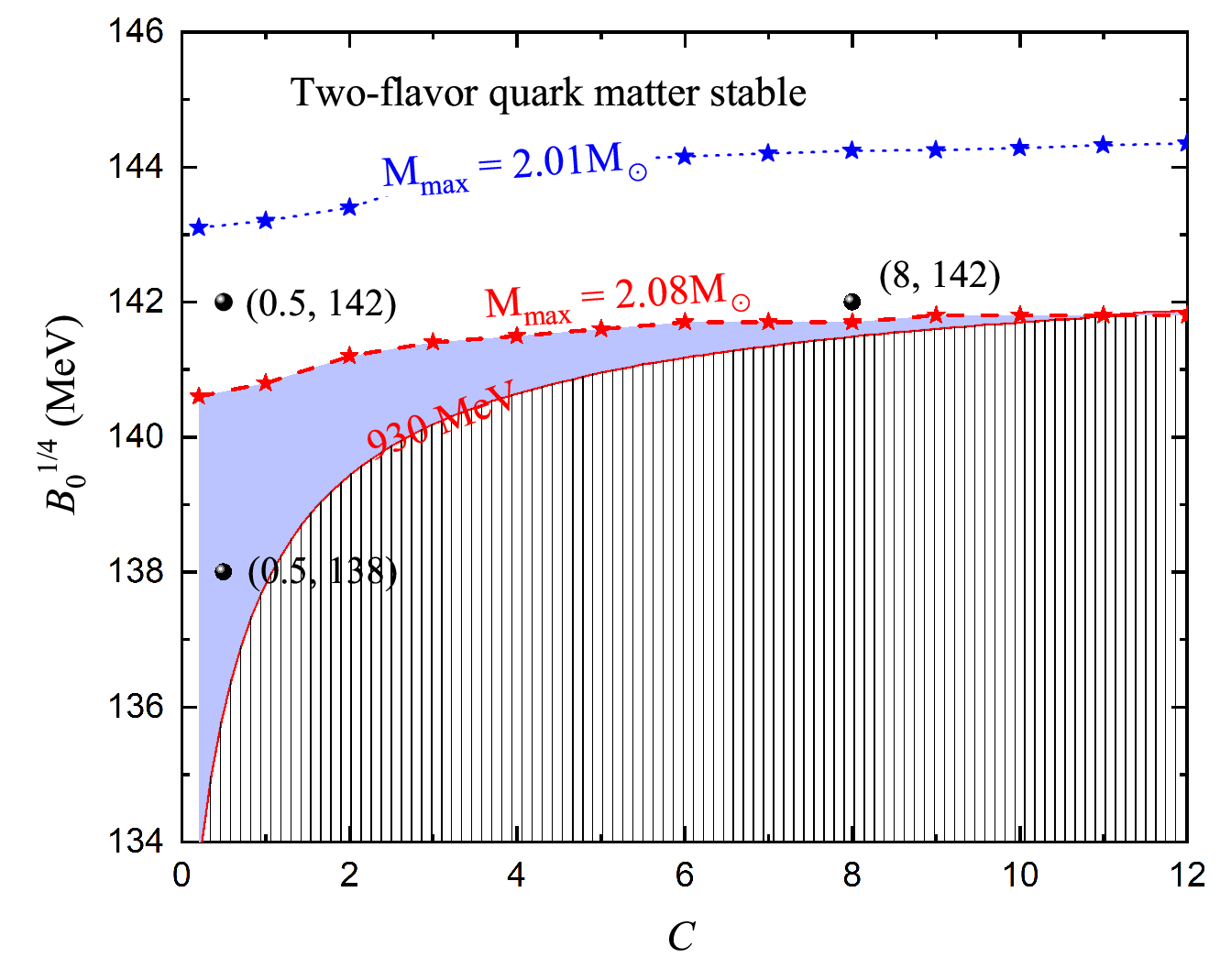}\\
  \caption{Stability windows for $ud$ quark matter in the $C-B_0^{1/4}$ plane. 
 The down-right black shaded area is forbidden, where the energy per baryon of $ud$ quark matter is less than 930 MeV. Meanwhile, 
 the region between the red dashed and blue dotted lines with star symbols 
  corresponds to the $ud$ quark matter stable region with an equation of state that can support an $ud$ quark star with the maximum mass in the range $2.01\leq M_{\text{max}}/M_{\odot}\leq 2.08$.  
  The selected typical model parameters are indicated
with solid black dots.
}\label{fig:C0B0}
\end{figure}

\subsection{Parameter space and equation of state of quark matter}

We note that the energy per baryon of $ud$ quark matter should be greater than 930 MeV in order not to contradict the standard nuclear physics requirement. 
Consequently, the stable condition and the two solar masses of compact stars usually set strict constraints on the parameter space of phenomenological models. 
In Fig.~\ref{fig:C0B0}, we show the parameter space for 
$ud$ quark matter in $C$ versus $B_0^{1/4}$ plane, where the 
region above the red solid line corresponds to the allowed region for the $ud$ quark matter. In contrast, the lower right shaded region below the red solid line are 
forbidden since the energy per baryon of $ud$ quark matter is less than 930 MeV. In addition, the measured gravitational mass of PSR J0740$+$6620~\cite{Fonseca-2021wxt} with a central value of $2.08~M_{\odot}$ and a lower limit of $2.01~M_{\odot}$ are also translated into the $C-B_0^{1/4}$ plane, indicated by the red dashed and blue dotted lines with star symbols.  
Therefore, only the region between the red solid and dashed blue lines is the stable region for $ud$ quark matter and in which the equation of state meets the requirement that the maximum mass of $ud$ quark stars can be as large as or even greater than $2.01M_{\odot}$, where $M_{\odot}$ is the solar mass.

\begin{figure}[h]
  \includegraphics[width=0.48\textwidth]{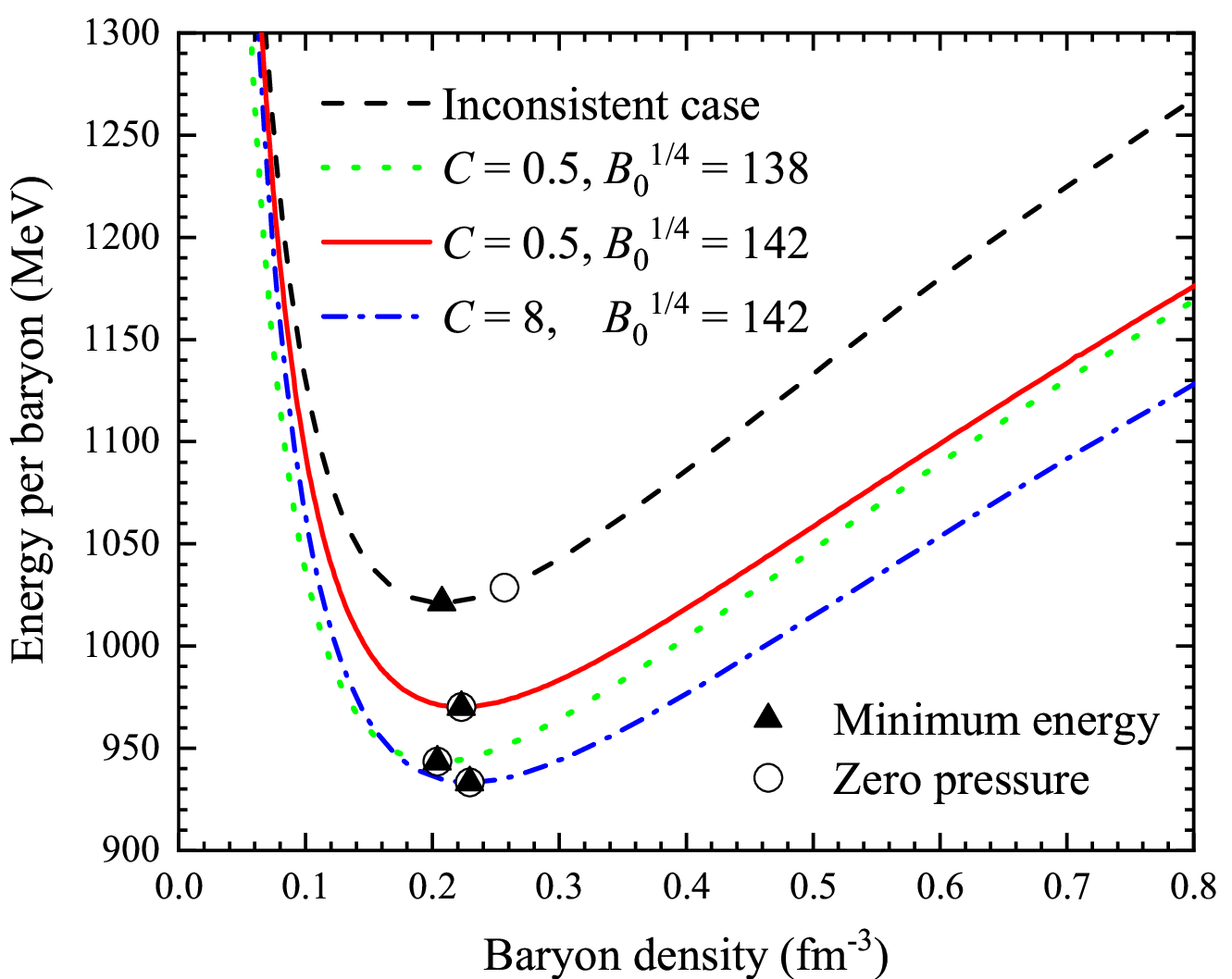}\\
  \caption{Density behavior of the energy per baryon of $ud$ quark matter with different values of $C$ and $B_0$. 
  The result from the model without adding 
  an effective bag constant is also shown for comparison. 
  It is clear that in the revised quasiparticle model, the zero pressure points denoted by circles are
exactly located at the minimum energy per baryon denoted by triangles, except for the inconsistent case denoted by the dashed line. }\label{fig:Enb}
\end{figure}

Considering the $ud$ quark matter with the equations of state satisfying the requirement of 
two solar masses of $ud$ quark stars, the parameters are set as $(C,~B_0^{1/4}/\text{MeV})=(0.5,~138),~(0.5,~142),~(8,~142)$, as indicated in  Fig.~\ref{fig:C0B0} by dots. 
Fig.~\ref{fig:Enb} shows
the energy per baryon of $ud$ quark matter as a function of the baryon density with the selected parameter sets.  
It can be seen that for the revised quasiparticle model the energy minima lie exactly at the corresponding zero pressure points, thus satisfying the thermodynamic requirement of the model. However, for the model without adding the effective bag constant as shown by the black dashed line, the energy minimum does not coincide with the zero pressure point. 
Furthermore, comparing the three typical lines, 
we find that 
with the increase in $B_0$ ($C$), the minimum energy per baryon increases (decreases). This is because, with increasing $B_0$ ($C$), the thermodynamic potential density of the system $\Omega$ in Eq.~(\ref{eq:Omega}) and in turn energy density increases (deceases) according to Eq.~(\ref{eq:E_P}).     

\begin{figure}[h]
  \includegraphics[width=0.48\textwidth]{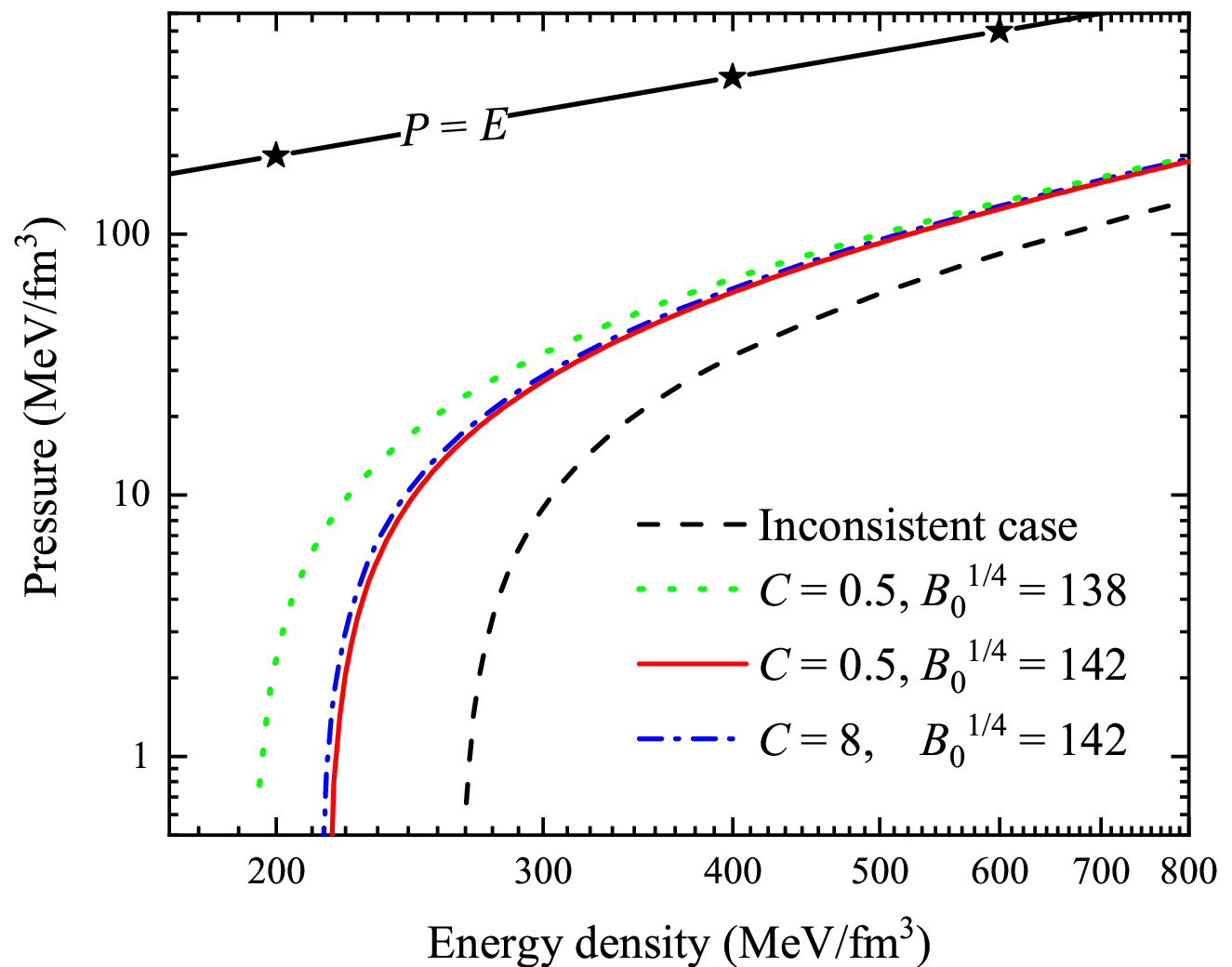}\\
  \caption{Equations of state of cold quark matter in the revised quasiparticle model with different values of $C$ and $B_0$. For comparison, we also show the results for inconsistent case and $P=E$. Conventions for line styles and colors are the same as in Fig.~\ref{fig:Enb}.  
  }\label{fig:EoS}
\end{figure}

In Fig.~\ref{fig:EoS}, we show the equations of state of $ud$ quark matter in the revised quasiparticle model, and the result for the inconsistent case and $P=E$ are also shown for comparison. From left to right, the green dotted, blue dash-dotted, red solid, and black dashed lines correspond to the revised quasiparticle model with parameter sets $(C,~B_0^{1/4}/\text{MeV})=(0.5,~138),~(8,~142),~(0.5,~142)$ and the inconsistent case, respectively. One can see that the pressure increases with the energy density without any exception. In particular, as the bag constant $B_0$ increases, the line moves to higher values of the energy density, whereas for a larger value of $C$, the line moves to a lower value of the energy density. Furthermore, except for the black dashed line (inconsistent case), the other three typical lines, representing the results obtained in the revised quasiparticle model, almost coincide 
at large energy densities.

\begin{figure}[h]
  \includegraphics[width=0.48\textwidth]{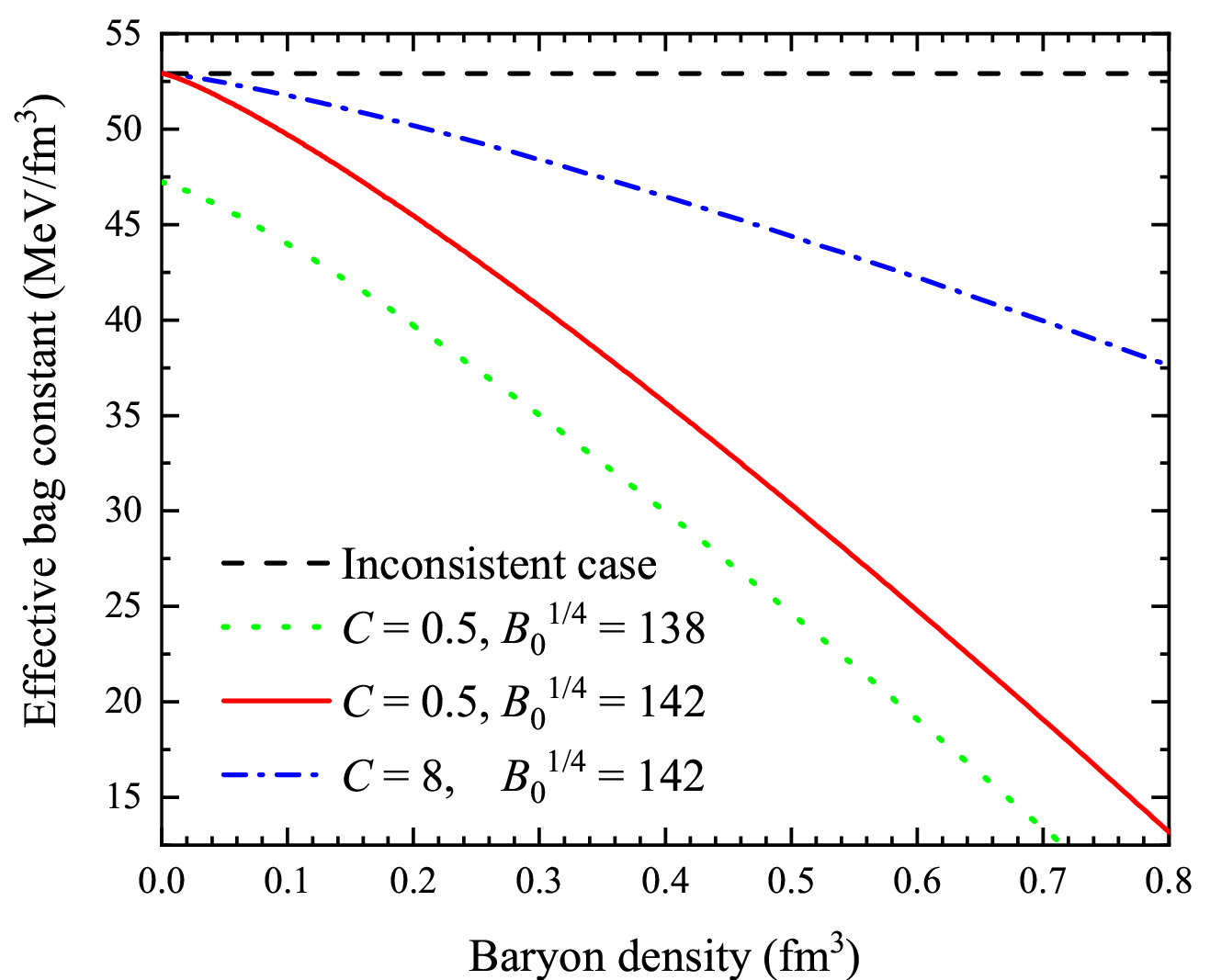}\\
  \caption{Density behavior of the effective bag constant with different values of $C$ and $B_0$. Conventions for line styles and colors are the same as in Fig.~\ref{fig:Enb}. 
  }\label{fig:Beffective}
\end{figure}

Fig.~\ref{fig:Beffective} shows the 
effective bag constant $B^*$ as a function of the baryon density for several parameter sets. For comparison, we also show results, indicated by the dashed line, for the inconsistent case without any correction to the energy density and pressure. 
For the revised quasiparticle model, as shown by the
solid, dotted, and dash-dotted lines, the values decrease
with increasing baryon density 
in all cases. This is consistent with the fact that color confinement becomes less important at sufficiently high densities.

\subsection{Mass-radius relations and tidal deformability of quark stars}

Quark stars are a class of hypothetical, theoretical compact stars composed of quark matter. Using the quark matter equation of state obtained from the revised two-flavor quasiparticle model shown in Eq.~(\ref{eq:E_P}), one can solve the Tolman-Oppenheimer-Volkoff (TOV) equations to obtain the mass-radius relation for dense quark stars. 
The equilibrium structure of a static spherically symmetric quark star is determined by the TOV equation
\begin{eqnarray}
\frac{\mathrm{d} P(r)}{\mathrm{d} r}=-\frac{G m E}{r^2} \frac{(1+P / E)\left(1+4 \pi r^3 P / m\right)}{1-2 G m / r},
\end{eqnarray}
and the subsidiary condition
\begin{eqnarray}
\frac{\mathrm{d} m(r)}{\mathrm{d} r}=4 \pi r^2 E,
\end{eqnarray}
where $G=6.7 \times 10^{-45}~ \mathrm{MeV}^{-2}$ is the gravitational constant, $r$ is the radial coordinate of the quark star, and $m$ is the gravitational mass contained within the radius $r$.

\begin{figure}[h]
  \includegraphics[width=0.48\textwidth]{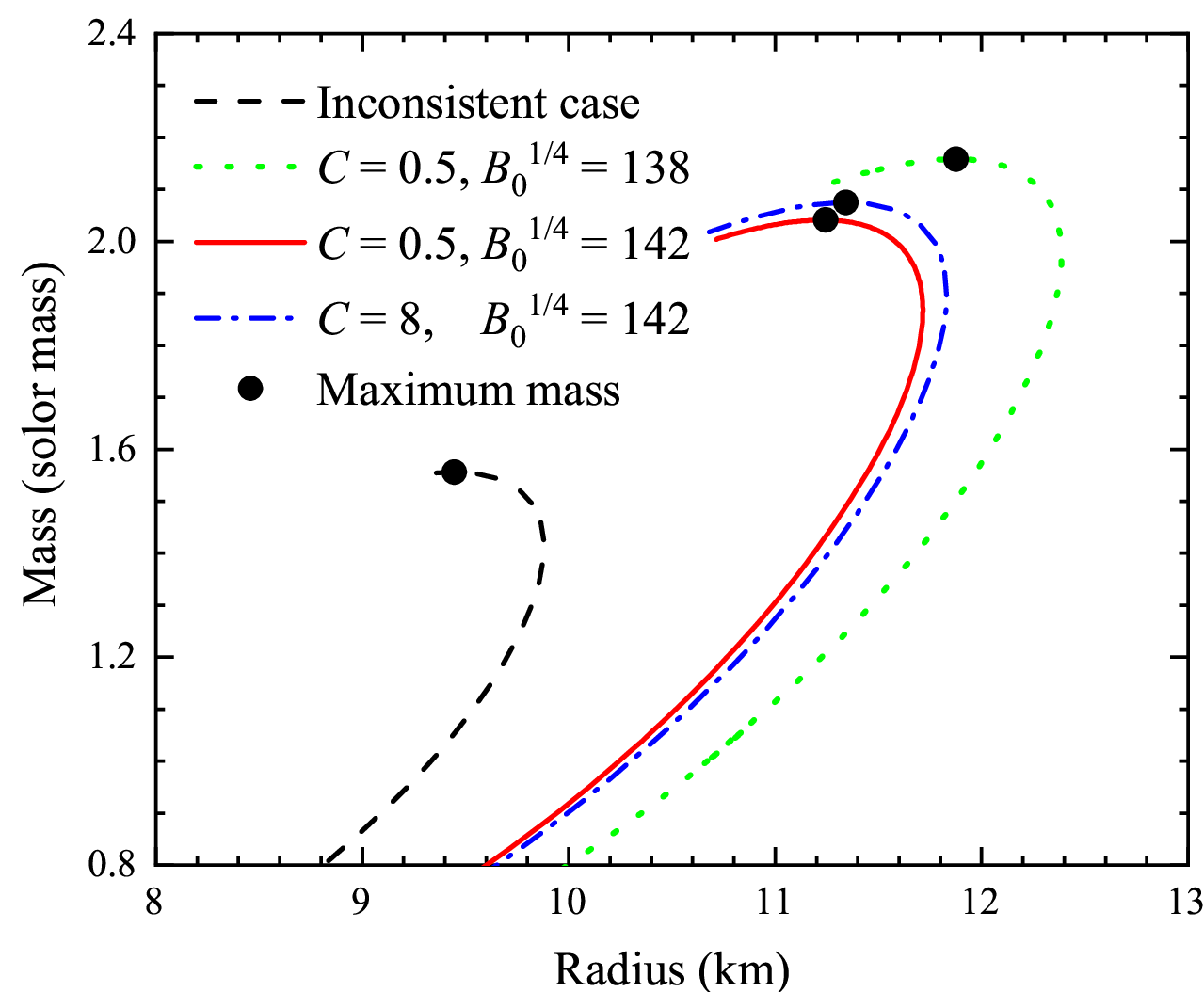}\\
  \caption{Mass-radius relations 
  for $ud$ quark stars. The maximum masses are denoted by the solid circles. Conventions for line styles and colors are the same as in Fig.~\ref{fig:Enb}. }\label{Fig:udMR}
\end{figure}

Taking the equation of state shown in Fig.~\ref{fig:Enb} as an input, the coupled TOV equations are solved numerically, and the obtained mass-radius relations of quark stars are shown in Fig.~\ref{Fig:udMR}. As can be seen from Fig.~\ref{Fig:udMR} that the inconsistent case supports a maximum mass of about 1.56 $M_{\odot}$ and the other curves show the maximum mass about two times the solar mass for the selected parameter sets in our model. 
Under the assumption that the $ud$ quark matter is more stable than the $uds$ quark matter, the observed two solar masses of compact stars~\cite{Demorest-2010bx,Antoniadis-2013pzd} can be described as a 2 $M_\odot$ $ud$ quark star.

\begin{figure}[h]
  \includegraphics[width=0.48\textwidth]{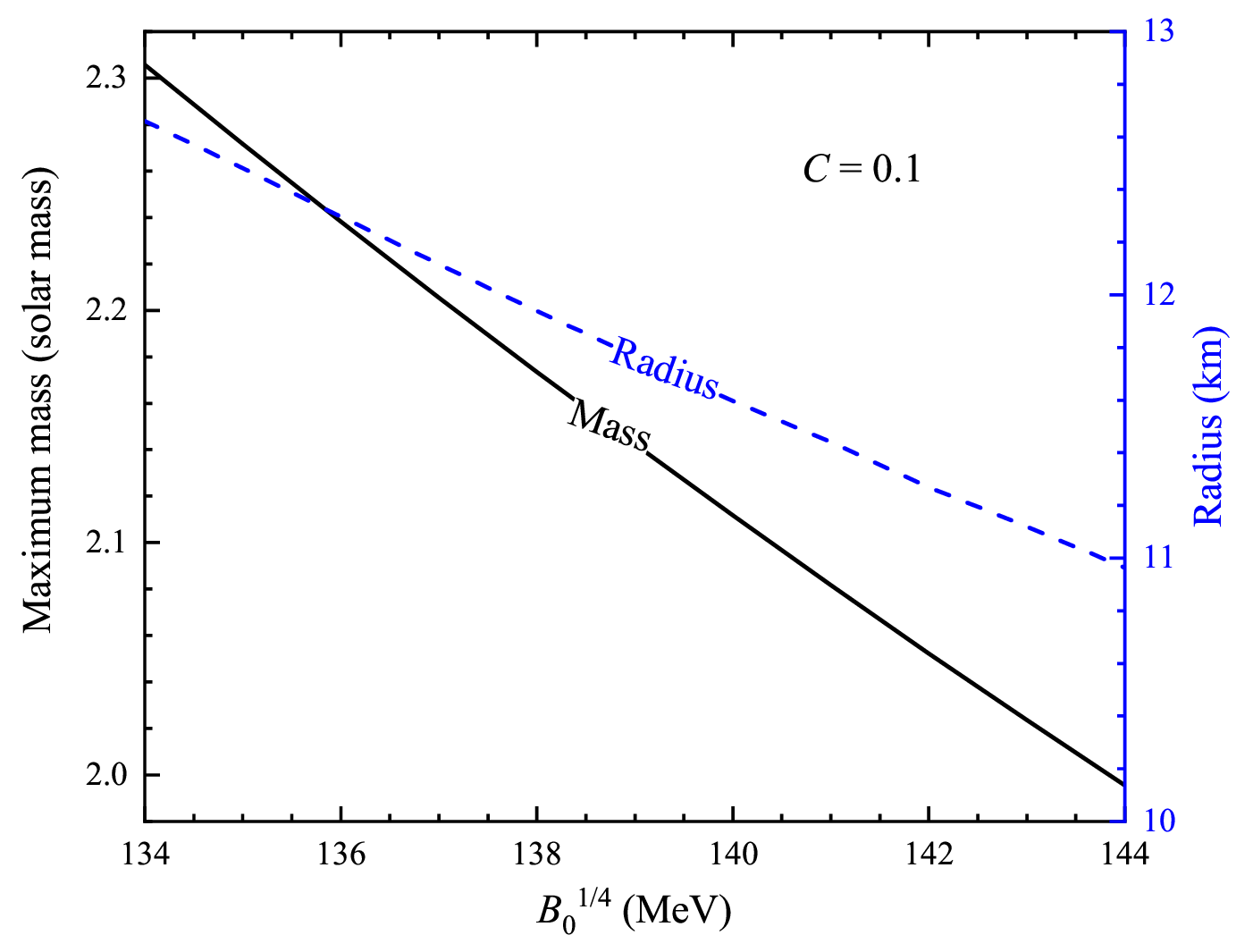}\\
  \caption{ Maximum mass and the corresponding radius of $ud$ quark stars as a function of the bag constant $B_0$ for $C=0.1$.  }\label{Fig:B0MR}
\end{figure}

Very recently, the determination of the gravitational mass of the compact
star PSR J0740+6620 has been updated to $2.08^{+0.07}_{-0.07}$~$M_{\odot}$~\cite{Fonseca-2021wxt}. %
Moreover, two more massive compact stars, MSR J0740+6620~\cite{NANOGrav-2019jur} and PSR J2215+5135~\cite{Linares-2018ppq}, have been measured to be $2.14 \pm_{0.09}^{0.10}~M_{\odot}$ and $2.27_{-0.15}^{+0.17}~ M_{\odot}$ respectively. 
Fig.~\ref{Fig:B0MR} shows the maximum mass and radius of $ud$ quark stars as functions of the bag constant $B_0^{1/4}$ at fixed $C=0.1$.  One can see that, both the maximum mass and the corresponding radius decrease with the increase in $C$. 
In particular, for the given equation of state with a small value of $C$, e.g. $C=0.1$, the maximum mass of a quark star composed of $ud$ quark matter has been calculated to be up to 2.31 $M_{\odot}$. This result demonstrates the potential for $ud$ quark stars to be more massive than neutron stars, which are limited to a maximum mass of about or greater than 2 $M_{\odot}$, suggesting that $ud$ quark stars may be an important component of the astrophysical landscape~\cite{Zhang-2019mqb}. 


We mention that, besides the mass-radius relation, owning to the recent detection of gravitational waves for event GW170817~\cite{LIGOScientific-2018cki} 
the measured tidal deformability also provides a stringent experimental constraints on the equation of state for compact stars~\cite{Drischler-2020fvz,Contrera-2022tqh,Yang-2021sqg}. Over the recent years, many efforts have been made to constrain the properties of quark star matter~\cite{Lourenco-2021lpn,Yang-2021bpe,Li-2021tur,Albino-2021zml,Pi-2022pjs,Yang-2019rxn,Xu-2022squ,Wang-2021jyo,Kumar-2022byc,Arbanil-2023yil}, based on the improved estimate of tidal deformability $\tilde{\Lambda}_{1.4}=190^{+390}_{-120}$ from the GW170817 event, where $\tilde{\Lambda}_{1.4}$ is the dimensionless
tidal deformability of a compact star with a mass of $1.4~M_{\odot}$. 
The response of the quark star to the gravitational field is described by the
tidal
Love number $k_2$, which depends on
the quark star structure and consequently on the mass and
the equation of state of quark matter. 
In General Relativity, the dimensionless tidal deformability $\tilde{\Lambda}$ is related to the $l=2$ tidal Love number $k_2$ as~\cite{Hinderer-2007mb,LIGOScientific-2018cki}  
\begin{eqnarray}
\tilde{\Lambda}=
\frac{2}{3\beta^5} k_2,
\end{eqnarray}
where $\beta=GM/R$ and $R$ are the compactness parameter and radius of the star, respectively. 
While the Love number $k_2$ can be expressed as~\cite{Flanagan-2007ix}  
\begin{eqnarray}
k_2&=& \frac{8 \beta^5}{5}(1-2 \beta)^2[2+2 \beta(y_R-1)-y_R] \nonumber\\
&& \times\{
4 \beta^3\left[13-11 y_R+\beta(3 y_R-2)+2 \beta^2(1+y_R)\right] \nonumber\\
&& +3(1-2 \beta)^2[2-y_R+2 \beta(y_R-1)] \ln (1-2 \beta) \nonumber\\
&&+2 \beta[6-3 y_R+3 \beta(5 y_R-8)] 
\},~~~~~
\end{eqnarray}
where 
$y_R\equiv y(R)$, and $y(r)$ is determined by solving the following differential equation
\begin{eqnarray}
r \frac{d y(r)}{d r}+y^2(r)+y(r) F(r)+r^2 Q(r)=0.
\end{eqnarray}
Here, $F(r)$ and $Q(r)$ are, respectively, defined as
\begin{eqnarray}
F(r) \equiv \frac{1-4 \pi r^2[E(r)-P(r)] G}{f(r)}
\end{eqnarray}
and
\begin{eqnarray}
Q(r) &\equiv & \frac{4 \pi}{f(r)}\bigg[5 E(r) G+9 P(r) G+\frac{E(r)+P(r)}{V_s^2(r)} G \nonumber\\
&& -\frac{6}{4 \pi r^2}\bigg]-4\left[\frac{m(r)+4 \pi r^3 P(r)}{r^2 f(r)} G\right]^2,
\end{eqnarray}
where $V_s^2$ represents the squared sound velocity of quark matter, and $f(r)$ takes the form $f(r)=1-2 Gm(r) / r$.   

\begin{figure}[h]
  \includegraphics[width=0.48\textwidth]{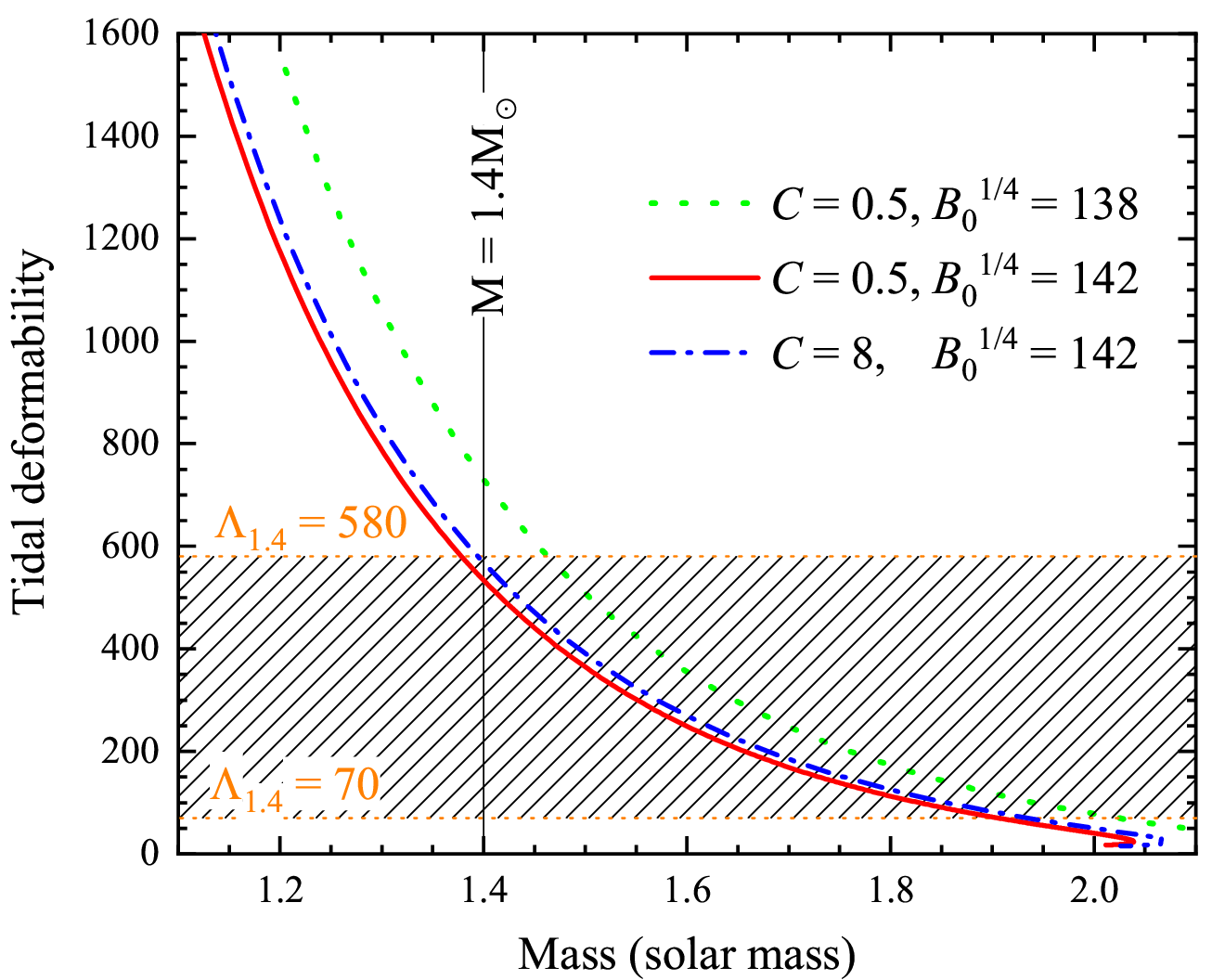}\\
  \caption{Tidal deformability of $ud$ quark stars as a function of the star mass for  different selected parameter sets. The black shaded area represents the improved estimate of the tidal deformability $\tilde{\Lambda}_{1.4}=190^{+390}_{-120}$ for GW170817. }\label{Fig:TidalD}
\end{figure}

Fig.~\ref{Fig:TidalD} shows the 
tidal deformability of $ud$ quark stars as a function of the star mass, for the revised quasiparticle model with different parameter sets in Fig.~\ref{fig:C0B0}. The vertical black solid line corresponds to the quark stars with a mass of $1.4 M_{\odot}$, whereas the black shaded area between the two horizontal orange dotted lines indicates the range of tidal deformability $\tilde{\Lambda}_{1.4}=190^{+390}_{-120}$, measured by LIGO and Virgo collaboration in the
event GW170817 
~\cite{LIGOScientific-2018cki}. 
Obviously, the tidal deformability decreases rapidly with increasing quark star mass, indicating that lighter quark stars can be deformed easier than heavier ones, as expected. 
Furthermore, as can be noticed from the plots, with the increase in $B_0$ the intersection of the green dashed line and the black solid line moves towards lower
values of tidal deformability. More specifically, except for the green dotted line, the intersections of red solid, blue dash-dotted lines with the vertical black solid line are located in the black shaded area. This indicates that the equations of state for these two selected parameter sets can simultaneously satisfy the constraint of the tidal deformability $70 \leq \tilde{\Lambda}_{1.4} \leq 580$ measured for GW170817 and the astrophysical observations of the compact star PSR J0740+6620 
with a mass of $2.08^{+0.07}_{-0.07}$~$M_{\odot}$. In particular, we have verified that for the parameter sets located in the blue shaded area shown in Fig.~\ref{fig:C0B0}, the corresponding equations of state can only satisfy the requirement that the massive compact stars with a mass about or even greater than two times the solar mass. While for the parameter sets taken from the region bounded by the red dashed line and the blue dotted line with star symbols can simultaneously satisfy the constraint of the tidal deformability $70 \leq \tilde{\Lambda}_{1.4} \leq 580$ extracted from GW170817 and the astrophysical observations of the compact stars PSR J0740+6620 and PSR J1614-2230 with a mass of $2.08^{+0.07}_{-0.07}$~$M_{\odot}$ and $1.97^{+0.04}_{-0.04}$~$M_{\odot}$, respectively.


We close this section by noticing that the recently observed massive compact stars have already ruled out a large number of equations of state since these equations of state are too soft to support a compact star with a mass about two times the solar mass. However, there are still various ways to obtain a massive quark star. One way is to assume the presence of the pion superfluid phase in the quark star~\cite{Mao-2014hga}, since the speed of sound in the pion superfluid phase is much larger than that of normal quark matter and ordinary nuclear matter~\cite{Lu-2021hvw}. In fact, the inclusion of non-Newtonian gravity effect can also significantly enhance the mass of quark stars~\cite{Lu-2017dsu,Yang-2021sqg}. 

\section{conclusions} \label{sec:conclusion}

The quasiparticle model is an essential phenomenological model for studying the properties of strongly interacting quark matter. However, the model with a medium-dependent quark mass that accounts for  
the strong interactions can lead to thermodynamic inconsistency problems. 
In particular, thermodynamic consistency limits the functional form of the renormalization subtraction on the chemical potentials. 
In this work, 
we have shown that the renormalization subtraction point should be taken as a function of the summation of the biquadratic chemical potentials if the quark's current masses vanish in order to ensure full thermodynamic consistency. 
Taking the simplest form, 
we have studied the thermodynamic properties of $ud$ quark matter and the mass-radius relation of
compact stars in a revised thermodynamically
consistent two-flavor quasiparticle model, considering the running of the QCD coupling
constant. 
It is found that there is allowed parameter space for $ud$ quark matter and simultaneously satisfies the recently observed massive pulsars with a mass larger than two times  the solar mass, reaching up to $2.13~M_{\odot}$.

We have also computed the dimensionless tidal deformability of $ud$ quark stars in the revised quasiparticle model. 
We found, however, that in order to further respect the upper limit of tidal deformability $\tilde{\Lambda}_{1.4}=190^{+390}_{-120}$ measured in the binary star merger GW170817 event, the maximum mass of an $ud$ quark star should not exceed $2.08~M_\odot$, i.e., $M_{\text{max}}\lesssim2.08~M_\odot$. 
Although the upper limit of the measured tidal deformability $\tilde{\Lambda}_{1.4}\leq 580$ extracted from the event GW170817 
poses an upper limit on the allowed maximum mass of compact stars, our results show that there is still a range of parameter space for the revised two-flavor quasiparticle model to simultaneously satisfy the stable condition, the astrophysical observations of the massive compact star with a mass about or slightly larger than two times the solar mass and the tidal deformability of quark stars in the range of $\tilde{\Lambda}_{1.4}=190^{+390}_{-120}$. In addition, it would be worth mentioning that if we relax the upper limit of the tidal deformability to $\tilde{\Lambda}_{1.4}\leq 800$~\cite{LIGOScientific-2017vwq}, the blue shaded area in Fig.~\ref{fig:C0B0} where the equations of state of $ud$ quark matter enable the support of the massive compact stars with maximum masses in the range $2.01\leq M_{\text{max}}/M_{\odot}\leq2.08$ and the tidal deformability $\tilde{\Lambda}_{1.4} \leq 800$ reported in Ref.~\cite{LIGOScientific-2017vwq}, would also be allowed.

Consistent with the previous studies using different methods~\cite{Ren-2020tll,Li-2022vof}, our results on the maximum masses of compact stars in this work have revealed that there is a range of parameters that can support the existence of $ud$ quark matter and are compatible with the observed massive pulsars with a mass about or greater than 
two solar masses. This has opened up new possibilities for understanding the physical properties of these incredibly dense stellar objects. It also provides insight into the potential implications of quark matter existing in the core of such massive pulsars.
In this paper, 
only the bulk properties of the $ud$ quark matter are considered. 
In fact, the finite size effect~\cite{Wen-2010zz,Wan-2020vaj} should be taken into account when the system is finite,  
such as quark matter nuggets~\cite{Xia-2022tvx,Xia-2018wdj}. 
These effects can significantly alter the physical properties of a system and can, therefore, have an important impact on the results of any simulations or experiments performed. For example, a finite system may exhibit unique behavior due to the presence of surface effects or the effects of a limited number of particles. 
Therefore, the extension of the model to include the surface and curvature terms 
would be highly relevant in the near future.

\section*{Acknowledgments}

ZYL thanks Prof.~Xin-Jian Wen for helpful and valuable discussions.  
This work is supported in part by 
the National Natural Science Foundation of China 
(Grant Nos.~12205093, 12204166, 12105097, 12005005, and 61973109), the Hunan Provincial Natural Science Foundation of China (Grant No.~2021JJ40188), 
the Scientific Research Fund of Hunan Provincial Education Department of China (Grant Nos.~19C0772 and 21A0297).

\bibliographystyle{aapmrev4-2}  
\bibliography{RefLuInsp}


\end{document}